\documentclass[conference]{IEEEtran}
\IEEEoverridecommandlockouts

\usepackage[table]{xcolor}
\usepackage[hyphens]{url}
\usepackage{hyperref}
\usepackage{tikz}
\usepackage{algorithmic}
\usetikzlibrary{arrows.meta, positioning}
\usetikzlibrary{shapes,arrows}
\usetikzlibrary{backgrounds}
\pgfdeclarelayer{background}
\pgfsetlayers{background}
\usepackage{pgfplots}
\usepackage{pgfplotstable} 
\usepackage{subcaption}
\usepgfplotslibrary{patchplots, colormaps, groupplots}
\pgfplotsset{compat=newest}
\usetikzlibrary{arrows, chains, positioning,fit,calc, shapes, shapes.geometric, matrix, external, spy}
\pgfplotsset{/pgfplots/colormap={winter}{rgb255=(0,0,255) rgb255=(0,255,128)}}
\pgfplotsset{/pgfplots/colormap={hsu}{cmyk=(0,1,.59,.26) cmyk=(1,.4, 0,.82)}}
\usepackage{tikz-uml}
\usepackage[english]{babel}
\usepackage{ifthen}
\usepackage{xstring}
\usepackage{calc}
\usepackage{pgfopts}
\usepackage[underline = false]{pgf-umlsd}
\usepackage[numbers, sort&compress]{natbib}
\usepackage[T1]{fontenc} 
\usepackage[cmintegrals]{newtxmath}
\usepackage{bm}
\usepackage{amsmath,amsfonts} 
\usepackage{mathtools}%
\usepackage{tabularx, booktabs}
\usepackage{multirow}
\usepackage{algorithmic}
\usepackage{graphicx}
\usepackage{textcomp}
\usepackage{xcolor}
\newcolumntype{C}[1]{>{\centering\arraybackslash}p{#1}}

\definecolor{hsurot}    {cmyk}{.00 1 .59 .26}
\definecolor{hsugrau}    {cmyk}{.38 .37 .39 .15}
\definecolor{hsugelb}    {cmyk}{0 .16 .80 0}
\definecolor{hsublau}    {cmyk}{1 .40 0 .82}
\definecolor{hsuturkis}    {cmyk}{1 .14 .60 .49}
\definecolor{hsugruen}    {cmyk}{.16 .16 .91 .28}
\definecolor{hsubraun}    {cmyk}{.00 .57 1 .17}
\definecolor{hsuorange}    {cmyk}{.01 .87 .77 .13}
\definecolor{lightlightgray}{cmyk}{0 0 0 .33}
\definecolor{dunkelgrau}{HTML}{5D6670}
\definecolor{babyblau}{HTML}{80B6F2}
\definecolor{hellblau}{HTML}{C7DAF0}
\definecolor{blaugrau}{HTML}{355170}
\definecolor{waldgrün}{HTML}{357529}
\definecolor{darkgray}{rgb}{0.66, 0.66, 0.66}
\definecolor{mygray}{RGB}{230, 230, 230} 
\definecolor{mycolor1}{HTML}{004c6d}
\definecolor{mycolor2}{HTML}{6ec0d0}
\definecolor{mycolor3}{HTML}{ef5675}
\definecolor{mycolor4}{HTML}{ffa600}
\definecolor{britishracinggreen}{rgb}{0.0, 0.26, 0.15}
\definecolor{dimgray}{rgb}{0.41, 0.41, 0.41}
\definecolor{feldgrau}{rgb}{0.3, 0.36, 0.33}
\definecolor{mediumtealblue}{rgb}{0.0, 0.33, 0.71}

\usepackage[printonlyused, nohyperlinks]{acronym}
\usepackage{hyperref}
\hypersetup{colorlinks=true, unicode=true, linkcolor=[rgb]{0.10,0.05,0.67}, citecolor=[rgb]{0.10,0.05,0.67}, filecolor=[rgb]{0.10,0.05,0.67}, urlcolor=[rgb]{0.10,0.05,0.67}}
\addto\extrasenglish{}
\addto\extrasenglish{}
\addto\extrasenglish{}
\addto\extrasenglish{}
\addto\extrasenglish{}
\addto\extrasenglish{}
\usepackage{enumitem}
\usepackage{rotating}

\DeclareMathOperator*{\argmin}{arg\,min}

\usepackage{blindtext}
\newlist{req}{enumerate}{2}
\setlist[req,1]{label=\textbf{R\arabic* -},ref=R\arabic*,align=left,leftmargin=*,widest=REQ99,labelsep=-2.3em} 
\setlist[req,2]{label=(\alph*),ref=\thereqi(\alph*),align=left,leftmargin=*,labelsep=-2.3em} 

\newlist{questions}{enumerate}{2}
\setlist[questions,1]{label=\textbf{RQ\arabic*},ref=RQ\arabic*,align=left,leftmargin=*,widest=RQ99,labelsep=-1.7em} %
\setlist[questions,2]{label=(\alph*),ref=\thequestionsi(\alph*),align=left,leftmargin=*,labelsep=-2.3em} %

\usepackage{booktabs}
\usepackage{siunitx}
\DeclareRobustCommand{\fulfill}[1]{%
  \begin{tikzpicture}[scale=0.12, baseline=-0.5ex]
    \draw (0,0) circle (1);
    \ifnum#1>0 \fill[fill opacity=1,fill=black] (0,0) -- (90:1) arc (90:90-#1*3.6:1) -- cycle;\fi
  \end{tikzpicture}%
}

\tikzset{
  basic/.style  = {draw, thin, text width=2cm, drop shadow, rectangle},
  root/.style   = {basic, rounded corners=2pt, thin, align=center,
                   fill=green!30},
  level 2/.style = {basic, rounded corners=6pt, thin,align=center, fill=green!60,
                   text width=2em},
  level 3/.style = {basic, thin, align=left, fill=pink!60, text width=6.5em},
  reqbox/.style  = {basic, rounded corners=3pt, thin, align=left, fill=white},
  funcbox/.style = {basic, rounded corners=3pt, thin, align=left, fill=lightgray},
  legreqbox/.style = {draw, drop shadow, rectangle, rounded corners=3pt, thin, align=center, fill=white},
  legfuncbox/.style = {draw, drop shadow,  rectangle, rounded corners=3pt, thin, align=center, fill=lightgray},
    line/.style = {draw},
}

\tikzset{
 decision/.style  = {diamond, black, draw, node distance=0.4cm},
  process/.style    = {fill=black!5,draw, thick, rounded corners=0.2cm, minimum height = 3em,
  minimum width = 3em, text width=4.2cm, align=center, inner xsep=0.2em, inner ysep=0.2em, node distance=0.4cm},
  object/.style    = {fill=black!5,draw, thick, rectangle, minimum height = 3em,
    minimum width = 3em, align=center, inner sep=1em,node distance=4em},
  pin/.style    = {fill=black!5,draw, thick, rectangle, minimum height = 0.6em,
    minimum width = 0.6em, node distance=-1pt, inner sep=0,font=\relsize{-3.5}},
  start/.style      = {fill=black,draw,circle,node distance=0.5cm}, 
  end/.style = {draw, circle, fill=white, node distance=0.5cm},
  endcenter/.style = {draw, circle, fill=black, minimum size=1.5em, inner sep=0pt, node distance=4em},
  group/.style      = {color=black,thin,rounded corners=0.8em, rectangle}, 
  groupCaption/.style      = {above=0.2cm,right=0.2cm,fill=white}, 
  input/.style    = {coordinate,node distance=2em}, 
  output/.style   = {coordinate,node distance=2em}, 
  between/.style args={#1 and #2}{ 
    at = ($(#1)!0.5!(#2)$)
  }  
}

\newcommand{\makeGroup}[5]{
  \draw [group](#2-0.5,#3+1.5-#5)rectangle(#2+#4+0.5,#3+1.5);
  \node at (#2,#3+1.5) [groupCaption] {\textsc{#1}};
}

\tikzumlset{
   fill usecase=white,
   usecase/style/.style={text width=2.5cm}
   }

\begin{document}
\bstctlcite{IEEEexample:BSTcontrol}
\title{Cost Optimized Scheduling in Modular Electrolysis Plants\\
\thanks{The authors gratefully acknowledge the funding of the project eModule (Support code: 03HY116) by the German Federal Ministry of Education and Research, based on a resolution of the German Bundestag.}}
\author{\IEEEauthorblockN{Vincent Henkel, Maximilian Kilthau, Felix Gehlhoff, Lukas Wagner, and Alexander Fay}
\IEEEauthorblockA{\textit{Institute of Automation Technology} \\
\textit{Helmut Schmidt University / University of the Federal Armed Forces Hamburg}, Germany \\
\{vincent.henkel, maximilian.kilthau, felix.gehlhoff, lukas.wagner, alexander.fay\}@hsu-hh.de}
}

\maketitle
\begin{abstract}
In response to the global shift towards renewable energy resources, the production of green hydrogen through electrolysis is emerging as a promising solution. Modular electrolysis plants, designed for flexibility and scalability, offer a dynamic response to the increasing demand for hydrogen while accommodating the fluctuations inherent in renewable energy sources. However, optimizing their operation is challenging, especially when a large number of electrolysis modules needs to be coordinated, each with potentially different characteristics.\\
To address these challenges, this paper presents a decentralized scheduling model to optimize the operation of modular electrolysis plants using the \textit{Alternating Direction Method of Multipliers}. The model aims to balance hydrogen production with fluctuating demand, to minimize the \textit{marginal Levelized Cost of Hydrogen} (mLCOH), and to ensure adaptability to operational disturbances. A case study validates the accuracy of the model in calculating mLCOH values under nominal load conditions and demonstrates its responsiveness to dynamic changes, such as electrolyzer module malfunctions and scale-up scenarios.
\end{abstract}

\setlength{\intextsep}{4pt}%

\vspace{-0.15cm}
\section{Introduction and Motivation}\label{:secIntroduction}
\vspace{-0.1cm}
Decarbonization is one of the greatest challenges of our time to mitigate climate change and secure a sustainable future. Hydrogen technologies play a crucial role in accelerating the decarbonization process. In particular, green hydrogen produced from renewable energy sources is a promising option to support this goal, as hydrogen serves as a high-density energy source for production facilities and solutions for transportation or storage. However, realizing the full potential of green hydrogen production through electrolysis is challenging due to the volatility and fluctuation of renewable energy sources. In addition, scaling hydrogen production to meet demand and production capacity is a significant obstacle.\cite{OUN+22}

In this context, modular electrolysis plants are an option for better adaptation to the variability of electricity generation from renewable energy sources and the increase in hydrogen demand. These plants allow the desired process to be configured by combining individual modules called \textit{Process Equipment Assemblies} (PEAs) \cite{BBE+22}. Such a modular structure increases the flexibility in the design of electrolysis plants and enables the seamless addition, removal, or replacement of electrolyzers. This, in turn, enhances the scalability of electrolysis capacities to meet evolving needs \cite{LKL+23}.

Due to the intrinsic capacity limitations of electrolysis-PEAs, for example due to physical and technical limitations, the construction of large-scale electrolysis plants requires the combination of a significant quantity of PEAs \cite{LKL+23}. Consequently, optimizing their operation and ensuring effective coordination can be challenging, especially when dealing with fluctuating operating conditions due to renewable energy sources or the heterogeneity of electrolysis-PEAs, each of which has unique operating characteristics \cite{LKL+23c}. Traditional centralized control approaches lack in scalability and flexibility to handle these fluctuating operating conditions \cite{nedic2018network}.

Thus, a decentralized \textit{Multi-Agent System} (MAS) for the control of modular electrolysis plants has been introduced in the study in \cite{SHM+23}. The MAS addresses the preceding challenges, and ensures adaptability, scalability, and robustness appropriate for modular electrolysis plants \cite{SHM+23}. As outlined in \cite{HSW24}, MAS are highly suitable for the integration into modular production. Furthermore, an analysis of agents' applications showed that they are often employed for control and scheduling \cite{RWG+23}. 

Building upon the work in \cite{SHM+23}, this paper introduces a scalable and robust decentralized optimization approach for cost-optimized scheduling in modular electrolysis plants by applying the \textit{Alternating Direction Method of Multipliers} (ADMM) in combination with MAS.

The contributions of this work are the following:
\begin{itemize}
    \item scalable, robust ADMM optimization algorithm capable of adapting to the number of electrolysis-PEAs and effectively responding to electrolyzer malfunctions
    \item implementation of this algorithm using a MAS for optimized decentralized control of a modular electrolysis plant
\end{itemize}

The remainder of this work is structured as follows: \autoref{sec:stateoftheart} provides an overview of prior and related work. The decentralized scheduling model and its implementation in MAS is explained in \autoref{sec:SchedulingModel}. The evaluation of the concept and discussion utilizing a case study are detailed in \autoref{sec:CaseStudy}. Finally, \autoref{sec:conclusion} consolidates the findings and outlines future directions for research.

\section{State of the Art}\label{sec:stateoftheart}

\subsection{Technical Background}\label{subsec:PreliminaryWork}

For flexible modular plants within the process industry, the \textit{Module Type Package} (MTP) \cite{2658} is an established modular automation concept that is characterized by low-effort, standardized integration and control of PEAs \cite{BBE+22}. Within this framework, each PEA is managed by its dedicated controller, which provides a standardized interface according to MTP guidelines \cite{2658}. This ensures efficient coordination of PEAs within modular plants by seamlessly integrating these PEAs into a higher-level control system known as the \textit{Process Orchestration Layer} (POL) \cite{2658}. As shown in \cite{BBE+22}, the MTP concept is also suitable for the application in modular electrolysis plants. 

\begin{figure}[h]
\centering
\resizebox{7cm}{!}{\input{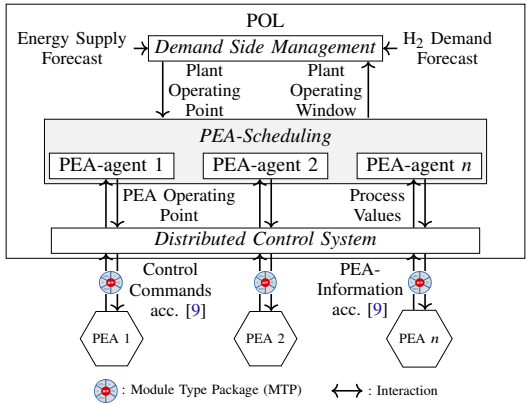}}
\captionsetup{font=small}
\caption{POL Refinement Architecture (Adapted from \cite{SHM+23})}
\label{fig:Architecture}
\end{figure}

However, the dynamics introduced by fluctuations in renewable energy sources and electricity prices pose novel challenges to the orchestration of modular plants that cannot yet be addressed with existing centralized POLs \cite{SHM+23}. To address these challenges, an initial approach for a domain-specific refinement of a POL architecture is presented in \cite{SHM+23}. This refinement involves the integration of a \textit{Demand Side Management} (DSM) component and a \textit{PEA-Scheduling} component (see \autoref{fig:Architecture}). The DSM is responsible for determining a plant operating point corresponding to the amount of hydrogen to be produced per time interval. The PEA-Scheduling is supposed to determine the scheduling (load distribution) of the electrolysis-PEAs, ensuring that the setpoint of the plant is realized in a cost-optimal way.

As depicted in \autoref{fig:Architecture}, each electrolysis-PEA is represented by a corresponding software agent, called PEA-agent. This MAS architecture allows for the scalable expansion of plant capacity by incrementally adding PEAs and their corresponding PEA-agents, commonly referred to as numbering-up \cite{2658}. 

The main objective of the MAS is to efficiently allocate the specified plant operating point from the DSM to the electrolysis-PEAs. Each PEA-agent is tasked with determining a specific amount of hydrogen to be produced by the represented electrolysis-PEA at a specific cost. Subsequently, the optimized load distribution is communicated to the \textit{Distributed Control System}, which then forwards control commands to the electrolysis-PEAs via the MTP interface. The PEA-Scheduling also aggregates the current availability of the electrolysis-PEAs into a plant operating window and communicates this information to the DSM.

Given the potential of the MTP to facilitate the seamless integration of diverse electrolysis-PEAs, it becomes essential to establish a foundational mathematical model within the MAS that accurately captures the heterogeneity in hydrogen production costs influenced by the various operational characteristics of different electrolysis-PEAs. As detailed in \cite{SHM+23}, the \textit{marginal Levelized Cost of Hydrogen} (mLCOH), which comprehensively accounts for cost components including \textit{Capital Expenditures} (CapEx), \textit{Operational Expenditures} (OpEx), \textit{Operation \& Maintenance} (O\&M), and efficiency over the operational lifetime \cite{GVE+22}, proves to be highly suitable for this purpose. The calculation of mLCOH is detailed in \autoref{sec:SchedulingModel}.

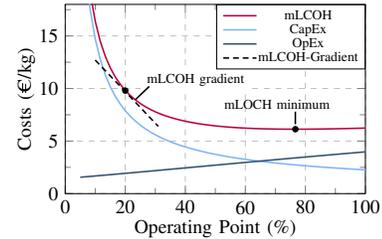
\begin{figure}[!h]
\centering
\resizebox{0.28\textwidth}{!}{\begin{tikzpicture}
\begin{axis} [ /pgf/number format/.cd,
        1000 sep={.},
        legend style={at={(1,1)}, anchor=north east, nodes={inner sep=0.5pt},
        font=\scriptsize, 
        },
        height=5cm, 
        width=7cm,
        xmin=0, xmax=100, minor x tick num=1, minor y tick num=1,
        ymin=0, ymax=18, 
        xlabel={Operating Point (\%)},
        ylabel={Costs (€/kg)},
        ylabel style={yshift=-5pt},
        xlabel style={yshift=4pt},
        ymajorgrids=true,
        xmajorgrids=true,
        grid style=dashed, 
        x tick label style={
            /pgf/number format/.cd,
            fixed,
            fixed zerofill,
            precision=0,
        },
]
\addplot[mark=,color=hsurot, thick] plot coordinates {
(5,29.8960916503602)
(6.66666666666667,23.1274599799534)
(8.33333333333333,19.0812643666489)
(10,16.3963961191092)
(11.6666666666667,14.4895169205251)
(13.3333333333333,13.0689523904419)
(15,11.9726566053683)
(16.6666666666667,11.1033979070201)
(18.3333333333333,10.399297906067)
(20,9.81910098400655)
(21.6666666666667,9.33424248648012)
(23.3333333333333,8.92431626231135)
(25,8.57435518692875)
(26.6666666666667,8.27313148950503)
(28.3333333333333,8.01205696463536)
(30,7.78444978012303)
(31.6666666666667,7.58503281914844)
(33.3333333333333,7.4095825197875)
(35,7.25467804609612)
(36.6666666666667,7.11751886708079)
(38.3333333333333,6.99578992376383)
(40,6.88756050448222)
(41.6666666666667,6.79120739011483)
(43.3333333333333,6.70535573502635)
(45,6.62883308557929)
(46.6666666666667,6.56063325182197)
(48.3333333333333,6.49988765399966)
(50,6.44584239976353)
(51.6666666666667,6.39783979804879)
(53.3333333333333,6.35530333910074)
(55,6.31772540540509)
(56.6666666666667,6.28465715127774)
(58.3333333333333,6.25570011738193)
(60,6.23049924282592)
(61.6666666666667,6.20873701043399)
(63.3333333333333,6.19012851644838)
(65,6.17441729873857)
(66.6666666666667,6.16137179077919)
(68.3333333333333,6.15078229455801)
(70,6.1424583859256)
(71.6666666666667,6.13622668198927)
(73.3333333333333,6.13192891295331)
(75,6.12942025104708)
(76.6666666666667,6.12856785741914)
(78.3333333333333,6.12924961453421)
(80,6.13135301702044)
(81.6666666666667,6.13477419833142)
(83.3333333333333,6.13941707420865)
(85,6.14519258691319)
(86.6666666666667,6.15201803666133)
(88.3333333333333,6.15981648874677)
(90,6.16851624653813)
(91.6666666666667,6.17805038196769)
(93.3333333333333,6.1883563163248)
(95,6.19937544517637)
(96.6666666666667,6.21105280208872)
(98.3333333333333,6.22333675654711)
(100,6.23617874208342)
};

\addplot[mark=,color=babyblau, thick] plot coordinates {
(5,28.3391182521774)
(6.66666666666667,21.5298588339988)
(8.33333333333333,17.4427954695917)
(10,14.7168301320704)
(11.6666666666667,12.7686350956267)
(13.3333333333333,11.3065464973564)
(15,10.1685288578685)
(16.6666666666667,9.25736088952817)
(18.3333333333333,8.51117450020431)
(20,7.8887242951428)
(21.6666666666667,7.36145577028308)
(23.3333333333333,6.90897285129621)
(25,6.51631841700762)
(26.6666666666667,6.17227462653637)
(28.3333333333333,5.86826313097353)
(30,5.59761188116769)
(31.6666666666667,5.35505346989407)
(33.3333333333333,5.13637397137282)
(35,4.93816211235362)
(36.6666666666667,4.75762685108618)
(38.3333333333333,4.5924625443855)
(40,4.44074782293071)
(41.6666666666667,4.30086873649233)
(43.3333333333333,4.17145963487615)
(45,4.05135718643525)
(46.6666666666667,3.939564249758)
(48.3333333333333,3.83522122017919)
(50,3.73758310698899)
(51.6666666666667,3.64600104731107)
(53.3333333333333,3.55990728612865)
(55,3.47880288721488)
(56.6666666666667,3.40224761272252)
(58.3333333333333,3.32985153770116)
(60,3.26126806219489)
(61.6666666666667,3.19618805651329)
(63.3333333333333,3.13433493093337)
(65,3.07546046390883)
(66.6666666666667,3.01934125604803)
(68.3333333333333,2.96577570302192)
(70,2.91458140091372)
(71.6666666666667,2.86559291361289)
(73.3333333333333,2.81865984465529)
(75,2.77364516615137)
(76.6666666666667,2.73042376568023)
(78.3333333333333,2.6888811786866)
(80,2.64891247932813)
(81.6666666666667,2.61042130713736)
(83.3333333333333,2.57331901048423)
(85,2.5375238908077)
(86.6666666666667,2.50296053405142)
(88.3333333333333,2.46955921778596)
(90,2.43725538420626)
(91.6666666666667,2.40598917062038)
(93.3333333333333,2.37570499024293)
(95,2.34635115711557)
(96.6666666666667,2.31787954982881)
(98.3333333333333,2.29024530944161)
(100,2.263406567609)

};

\addplot[mark=,color=blaugrau, thick] plot coordinates {
(5,1.55697339818283)
(6.66666666666667,1.59760114595459)
(8.33333333333333,1.63846889705714)
(10,1.67956598703881)
(11.6666666666667,1.72088182489847)
(13.3333333333333,1.76240589308551)
(15,1.80412774749984)
(16.6666666666667,1.84603701749193)
(18.3333333333333,1.88812340586273)
(20,1.93037668886376)
(21.6666666666667,1.97278671619704)
(23.3333333333333,2.01534341101514)
(25,2.05803676992113)
(26.6666666666667,2.10085686296865)
(28.3333333333333,2.14379383366183)
(30,2.18683789895534)
(31.6666666666667,2.22997934925438)
(33.3333333333333,2.27320854841468)
(35,2.3165159337425)
(36.6666666666667,2.35989201599461)
(38.3333333333333,2.40332737937833)
(40,2.44681268155151)
(41.6666666666667,2.4903386536225)
(43.3333333333333,2.5338961001502)
(45,2.57747589914404)
(46.6666666666667,2.62106900206397)
(48.3333333333333,2.66466643382047)
(50,2.70825929277454)
(51.6666666666667,2.75183875073772)
(53.3333333333333,2.79539605297208)
(55,2.83892251819021)
(56.6666666666667,2.88240953855522)
(58.3333333333333,2.92584857968077)
(60,2.96923118063103)
(61.6666666666667,3.0125489539207)
(63.3333333333333,3.05579358551502)
(65,3.09895683482974)
(66.6666666666667,3.14203053473116)
(68.3333333333333,3.18500659153609)
(70,3.22787698501187)
(71.6666666666667,3.27063376837638)
(73.3333333333333,3.31326906829802)
(75,3.35577508489571)
(76.6666666666667,3.39814409173891)
(78.3333333333333,3.44036843584761)
(80,3.48244053769231)
(81.6666666666667,3.52435289119406)
(83.3333333333333,3.56609806372442)
(85,3.6076686961055)
(86.6666666666667,3.64905750260991)
(88.3333333333333,3.69025727096081)
(90,3.73126086233187)
(91.6666666666667,3.77206121134732)
(93.3333333333333,3.81265132608187)
(95,3.8530242880608)
(96.6666666666667,3.8931732522599)
(98.3333333333333,3.9330914471055)
(100,3.97277217447443)
};

\addplot[mark=none, color=black, thick, densely dashed, domain=10:31] {15.7 - 0.3*x};

\addlegendentry{mLCOH}
\addlegendentry{CapEx}
\addlegendentry{OpEx}
\addlegendentry{mLCOH-Gradient};

\addplot[only marks, mark=*, mark size=1.4pt, mark options={draw=black}]
  coordinates {(20,9.8)};
\node[anchor=west, fill = white] at (axis cs:25,10.8) {\textcolor{black!100}{\scriptsize mLCOH gradient}};
\draw[-, line width=0.8pt] (axis cs: 28,10.2) -- (axis cs: 23.33,8.924);

\addplot[only marks, mark=*, mark size=1.4pt, mark options={draw=black}]
  coordinates {(76.6666666666667,6.12856785741914)};

\node[anchor=west] at (axis cs: 50.6666666666667,8.52856785741914) {\textcolor{black!100}{\scriptsize mLOCH minimum}};
\draw[-, line width=0.8pt] (axis cs: 76.6666666666667,7.8) -- (axis cs: 76.6666666666667,6.6);
\end{axis}
\end{tikzpicture}}
\vspace{-0.25cm}
\captionsetup{font=small}
 \caption{OpEx and CapEx contributions to mLCOH and the mLCOH gradient for a 10MW \textit{Proton Exchange Membrane} (PEM) electrolyzer at varying operating points with electricity cost of 0.06€/kWh (Adapted from \cite{GVE+22})}
\label{fig:mLCOHGradient}
\end{figure}

As shown in \autoref{fig:mLCOHGradient}, the mLCOH can be calculated as a function of the operating point of the electrolyzer. By determining the gradient at the current operating point, it is possible to estimate how a change in the operating point would affect the associated mLCOH value.

To achieve cost-optimal operation in a modular electrolysis plant, the PEA-Scheduling must fulfill the following requirements \cite{SHM+23}:
\begin{req}[itemindent=2.3em]
    \item{\textbf{Autonomous Planning of Operating States:}} 
     The system must autonomously plan the operating state (e.g., idle or production) of the electrolyzers, taking into consideration the technical limitations (e.g., start-up time) to ensure that they are operated cost-optimally and in accordance with the variable plant operating point (H$_2$-Demand). 
    \item{\textbf{Seamless Integration of Heterogeneous Electrolyzers:}}
    To provide the required scalability and adaptability of a modular electrolysis plant, the system must facilitate the low-effort integration of heterogeneous electrolyzers of different types, vendors, and characteristics, reducing the need for manual configuration steps.
    \item{\textbf{Resilience to Resource Failures:}}
    The system must prove its robustness by reacting to error states or malfunctions of a resource (e.g., electrolyzer), and ensure continous hydrogen production. 
%
     \item{\textbf{Decentralized Cost-Optimized Scheduling:}}
      Given the modular architecture of a modular electrolysis plant and the complexity and variability of the control parameters, which may not be effectively managed by centralized methods, the PEA scheduling must be decentralized. Thus, the system must perform decentralized scheduling among heterogeneous electrolyzers to meet variable hydrogen demand while minimizing production costs.  
%
%
    \item{\textbf{Efficient Operation Considering Degradation:}} The system must implement resource-efficient operation strategies to minimize frequent load changes and start-stop cycles, thereby mitigating potential electrolyzer degradation risks. 
     \item{\textbf{Electrolyzer Monitoring and Control:}} The system must monitor the current state of the electrolyzers and convert the optimized setpoints, derived from the cost-optimized scheduling, into actionable control commands. 
\end{req}
\vspace{-0.15cm}
\subsection{Related Work}\label{subsec:RelatedWorks}
\vspace{-0.15cm}
Numerous studies have addressed scheduling in electrolysis plants, broadly categorized into two approaches: optimization methods, such as \textit{Mixed Integer Linear Programming} (MILP) optimization \cite{VMZ21,VCF+22,FPB+21}, and heuristic optimization \cite{FaLi19,LKL+23b,LKL+23c,ZZT+22}. Agent-based approaches \cite{KGM+22,MPD+22,BTG+20} have found utility in similar applications. However, they are not exclusively focused on stand-alone electrolysis plants.
This section provides an overview of relevant research in this domain. The related works have been selected based on their significance and applicability to the topic under consideration.

In \cite{VMZ21}, a centralized scheduling model for alkaline water electrolysis is developed as a MILP model with decision variables representing operational states and transitions. The model incorporates operational characteristics as constraints, considering three operational states (production, standby, idle) and transitions (cold/full start, shutdown) for an accurate representation of electrolysis operations.
The model enables optimal sizing of electrolyzers and provides production schedules based on data sets of intermittent energy and electricity prices. However, the model does not consider error states as it solely focuses on the target behavior (R3). Additionally, it lacks provisions for translating optimized schedules into control commands or incorporating monitoring functionalities (R7). Moreover, its centralized nature poses a limitation (R4), hindering adaptability to the distributed and decentralized structure of modular electrolysis plants \cite{EHS15}.
%

The optimization in \cite{VCF+22} focuses on sizing and scheduling a power-to-hydrogen system connected to a hydrogen refueling station. The objectives are the identification of the optimal system dimensions and the scheduling of hydrogen production, accounting for various electricity supply scenarios (grid, solar, and hydro). The results offer valuable insights into the design and operational parameters with a focus on minimizing LCOH. Similar to \cite{VMZ21}, this study neglects error states (R3) and lacks integration for monitoring and translation into control commands (R7). Additionally, the model's centrality limits its adaptability to modular electrolysis plants (R4).

In \cite{FPB+21}, the authors present a real-time optimization control system for a PEM electrolyzer. Employing a MILP model that captures the nonlinear aspects of the electrolyzer, the optimization is applied every minute, aiming to meet hydrogen demand at a lower cost than achievable using a linearized model. The supervisory controller calculates the power value for the electrolyzer and transmits it to an intermediary controller, which subsequently forwards the command to the lower-level controller of the electrolyzer. However, the study misses the consideration of error states (R3) and limits the scope of optimization to a single PEM electrolyzer (R4).


The work presented in \cite{FaLi19} proposes an adaptive control strategy aiming to optimize the operation of alkaline electrolyzers. This strategy incorporates three control approaches: Simple start-stop, slow start, and segment start. The research highlights the efficiency of the segment start strategy in minimizing the frequency of start-up and shutdown cycles. However, the study does not consider costs (R4), degradation risks (R6), and lacks the integration with a controller (R7).

Based on \cite{FaLi19}, the approaches in \cite{LKL+23c} and \cite{LKL+23b} present an adaptive process control strategy. Utilizing a rule-based distribution method, these approaches efficiently allocate the load among stack units in modular electrolysis plants. However, they fall short in addressing costs (R4) and conducting malfunction analyses (R3).


The optimization in \cite{ZZT+22} focuses on hydrogen production from off-grid wind power sources by combining electrolyzer power allocation with wind power prediction. The control strategy allocates electrolyzers to specific operation modes for consistent operation. Nonetheless, the study misses considerations for costs (R4), malfunctions (R3), and lacks clarification on control and monitoring functionalities (R7).

The work in \cite{KGM+22} proposes a stochastic, agent-based model for planning a Multi-Energy-Microgrid, considering electricity, hydrogen, and gas energy sectors. The objective is to optimize cooperation among these sectors, addressing uncertainties in energy generation, storage, and demand. The model uses ADMM to describe interactions involving common variables and coupling constraints, and solves three correlated optimization problems. 
However, resource states are not sufficiently taken into account (R1), and although the robustness of the approach is partially considered by including stochastic influencing factors, fault states of resources are disregarded (R3). Additionally, no resource monitoring or control functions are provided (R7).

In \cite{MPD+22}, a cooperative planning and operation method for a wind-hydrogen-heat MAS system is presented. It addresses the challenges of balancing the interests of multiple participants in the system, achieving efficient energy usage by means of cooperation. The model uses ADMM to solve two sub-problems. The first focuses on minimizing planning and operating costs, and the other addresses the payment bargaining problem. However, the approach lacks sufficient details regarding the electrolyzer's operational states, and it does not consider degradation (R1, 6). While a sensitivity analysis of the wind power feed-in tariff partially addresses the robustness of the approach, explicit attention to fault states of resources is still absent (R3). Moreover, no resource monitoring or control functionalities are provided (R7).

In the study presented in \cite{BTG+20}, the authors introduce a decentralized multi-agent energy management system for a marine hydrogen energy system. This system optimizes energy distribution among components, including an electrolyzer, a fuel cell, and hydrogen storage. An electrolyzer agent manages and optimizes the electrolyzer's operations, while DSM is handled by a respective DSM agent. However, resource states are not considered (R1), and the work solely focuses on regular operation, although resource failures could potentially be detected by the provided monitoring (R3). Moreover, the absence of cost considerations poses a limitation (R4).

The overview of related works and requirements is shown in \autoref{tab:requirementsRelatedWorks}. 
The analysis of MILP optimization studies consistently reveals a predominant emphasis on optimization, often at the neglect of addressing connections to a controller for monitoring and control purposes. These algorithms are also centralized. Given the decentralized automation (both hardware and software) inherent in modular electrolysis plants, and given the fundamental importance of low-effort adaptability and scalability in such systems \cite{BBE+22, LKL+23c}, the scheduling model should be implemented in a decentralized manner \cite{SHM+23, EHS15, SANA22}. 
Nevertheless, integrating start-up costs to minimize degradation, as seen in studies like \cite{VMZ21, VCF+22}, is a positive aspect to consider in the model in this work.

\vspace{0.1cm}
\begin{table}[h!]
  \centering
  \captionsetup{font=small}
  \caption{Evaluation of scientific research. Semantics: \fulfill{100} fulfilled, \newline \fulfill{50} partially fulfilled, \fulfill{0} not fulfilled, [-] not evaluable}
 \renewcommand{\arraystretch}{1.1} %
  \begin{tabular}{|l|c|c|c|c|c|c|}
    \hline
     & \begin{sideways}\textbf{R1}\;\,\end{sideways} & \begin{sideways}\textbf{R2}\end{sideways} & \begin{sideways}\textbf{R3}\end{sideways} & \begin{sideways}\textbf{R4}\end{sideways} & \begin{sideways}\textbf{R5}\end{sideways} & \begin{sideways}\textbf{R6}\end{sideways}\\
    \hline
    \multicolumn{7}{|c|}{\cellcolor{mygray} MILP optimization} \\
    \hline
    \citet{VMZ21} & \fulfill{100} & [-] & \fulfill{0} & \fulfill{50} & \fulfill{100}  & \fulfill{0} \\
    \hline
    \citet{VCF+22} & \fulfill{100} & [-] & \fulfill{0}  & \fulfill{50}  & \fulfill{100} & \fulfill{0} \\  
    \hline
    \citet{FPB+21} & \fulfill{100} & [-] & \fulfill{0}  & \fulfill{50}  & \fulfill{50} & \fulfill{100}  \\
    \hline
    \multicolumn{7}{|c|}{\cellcolor{mygray} Heuristic optimization} \\
    \hline
    \citet{FaLi19} & \fulfill{50} & [-] & \fulfill{0}  & \fulfill{0} & \fulfill{0} & \fulfill{0} \\
    \hline
    \citet{LKL+23b, LKL+23c} & \fulfill{50} & \fulfill{50} & \fulfill{0}  & \fulfill{0}  & \fulfill{100} & [-] \\
    \hline
    \citet{ZZT+22} & \fulfill{50} & [-] & \fulfill{0}  & \fulfill{0} & \fulfill{100}& [-]  \\
    \hline
    \multicolumn{7}{|c|}{\cellcolor{mygray} Agent-based optimization} \\
     \hline
    \citet{KGM+22} & \fulfill{50} &  [-] & \fulfill{50}  & \fulfill{100}  & \fulfill{100} & \fulfill{0}  \\
     \hline
    \citet{MPD+22} & \fulfill{0} &  [-] & \fulfill{50}  & \fulfill{100} & \fulfill{0} & \fulfill{0}  \\
    \hline
    \citet{BTG+20} & \fulfill{0} & [-] & \fulfill{0}  & \fulfill{0}  & \fulfill{100}& \fulfill{100}  \\
    \hline
  \end{tabular}
  \label{tab:requirementsRelatedWorks}
\end{table}

Heuristic optimization approaches \cite{FaLi19,LKL+23b,LKL+23c,ZZT+22}, on the other hand, miss cost considerations, making them incapable of achieving economically optimized operation. Moreover, they lack provisions for establishing connections to controllers for monitoring or control purposes. Only \cite{LKL+23b, LKL+23c} explicitly discuss the low-effort integration and scale-up of electrolyzers within a modular electrolysis plant. However, the feasibility of this approach is not explained in detail in the case study, which underlines the need for a more in-depth investigation.\\
While agent-based approaches use a decentralized strategy suitable for modular electrolysis plants, they lack precision in modeling electrolyzer behavior and focusing mainly on target behavior. Despite recognizing agents as intrinsically robust \cite{RWG+23,HSW24}, their ability to respond to malfunctions is limited in these models. Nonetheless, \cite{KGM+22, MPD+22} emphasize the scalability and robustness of ADMM which can potentially also be used to react to malfunctions of electrolysis-PEAs in this work. 

Existing approaches focus on achieving target behavior and lack consideration of deviations, such as electrolysis-PEA malfunctions. However, modular electrolysis plants, with multiple electrolysis-PEAs \cite{SHM+23,LKL+23c}, require adaptable, resilient control and optimization systems. Latest research reveals a gap in addressing deviations and ensuring the robustness of control strategies. There is a need for comprehensive solutions that involve continuous monitoring to track the operational status capturing the real plant behavior \cite{FPB+21}, and the development of a decentralized optimization model tailored for modular electrolysis plants. This work aims to address this gap by introducing a novel scheduling model designed to address the unique challenges of modular electrolysis plants.


\vspace{-0.1cm}
\section{Decentralized ADMM Scheduling Model}\label{sec:SchedulingModel}
ADMM is a decentralized optimization algorithm designed to simultaneously solve distributed optimization problems with a shared objective. Previous research \cite{SANA22,KGM+22, MPD+22} has demonstrated the effectiveness of ADMM in addressing cooperative planning problems in the energy domain. It exhibits characteristics such as rapid convergence, robustness, scalability, and adaptability \cite{MPD+22,KGM+22}. The convergence guarantee for convex problems is proven in \cite{Boy10}.

Therefore, ADMM is used in the context of scheduling in modular electrolysis plants, where these characteristics are particularly important. In the following section, the basic principles of ADMM and the associated equations for cost-optimal scheduling based on mLCOH are explained. 
Importantly, the scheduling focuses on the core cost components and excludes certain additional cost factors such as water costs analogous to \cite{VMZ21,GVE+22,FPB+21,LKL+23c} due to the limited availability of robust data.

ADMM relies on the Lagrangian function $\mathcal{L}_\rho$ (see \autoref{eq:1}), which is composed of an objective function, denoted as $f(x)$, a set of constraint functions $g(z)$, a coupling term expressed as $y^T (Ax+Bz-c)$, with $A$ and $B$ representing the respective matrices related to certain system parameters and $c$ denoting a constant vector, and an associated but optional penalty term $\frac{p}{2} \cdot |Ax-b|^2 $ \cite{Boy10}. 
\vspace{-0.1cm}
\begin{align}
\begin{split}
\mathcal{L}_\rho(x,z,y) = & f(x) + g(z) + y^T(Ax + Bz - c) +\\
&  \frac{p}{2}\cdot |Ax-b|^2\label{eq:1}
\end{split}
\end{align}
Each PEA-agent representing an electrolysis-PEA aims to optimize its operation by minimizing its mLCOH, expressed in the objective function $f(x)$.
To avoid high frequency start-up and shutdown, which can lead to increased degradation of an electrolysis-PEA, start-up costs $C^{SU}$ are included \cite{VMZ21,VCF+22}. These costs are multiplied by the binary variable $Y_t$, set to 1 according to \autoref{eq:xyz} during electrolysis-PEA start-up. The variable mH2$_t$ represents the hydrogen production rate in kg/h and is defined by \autoref{eq:6}. To prevent division by zero when the electrolyis-PEA is in the idle state, a small value $\delta$ is introduced.
\vspace{-0.2cm}
\begin{align}
f(x) &= \text{O\&M}_{t} + \frac{\text{CapEx}_{t} + \text{OpEx}_{t}(x) + C^{SU}\cdot Y_t}{\text{mH2}_{t}(x) + \delta} \label{eq:2}
\end{align}
In this context, CapEx$_t$ represents the annuity in € in period~$t$, calculated in \autoref{eq:xx} as described in \cite{KSF+13}, where $UT$ defines the utilization time of the electrolysis-PEA in years and $r$ specifies the discount rate in \%.
\vspace{-0.2cm}
\begin{align}
    \text{CapEx}_{t} &= \frac{\text{CapEx}_{0} \cdot r \cdot (1+r)^{UT}}{(1+r)^{UT} - 1}\label{eq:xx}
\end{align}
The fixed costs of O\&M$_t$ per period, as defined in \autoref{eq:xxx}, are calculated using a methodology analogous to \cite{BRP22}. It is directly proportional to the initial CapEx$_0$, with a scaling factor denoted as \textit{OMF}. This factor is influenced by both the production quantity under nominal power
(\text{mH2}$_{\text{nom}}$) and the load factor (\textit{LF}), which indicates the percentage of full load hours per year \cite{BRP22}.
\vspace{-0.2cm}
\begin{align}
    \text{O\&M}_{t} &= \frac{\text{CapEx}_{0} \cdot \text{\textit{OMF}}}{\text{\textit{LF}} \cdot 8760 \text{(hrs/year)} \cdot \text{mH2}_{\text{nom}}}\label{eq:xxx}
\end{align}
In \autoref{eq:3}, the operating costs OpEx$_t$ arise from the operational setpoint Op$_t$ of the electrolysis-PEA in period $t$ in \%, multiplied by the capacity P${_\text{el}}$ in kW, the electricity cost C${_{\text{E},t}}$ in €/kWh, and the scheduling interval $\Delta_{\text{int}}$ in h. Following \cite{BBS+23}, OpEx are considered only when the electrolysis-PEA is actively operational, indicated by the binary variable $x_{\text{operation},t}$ with a value of 1, as detailed in \autoref{eq:xxxx}.
\vspace{-0.1cm}
\begin{align}
    \text{OpEx}_{t} &= \left(\text{Op}_{t}(x) \cdot \text{P}_{\text{el}} \cdot \text{C}_{\text{E},t} \cdot \Delta_{\text{int}} \right) \cdot x_{\text{operation},t} \label{eq:3}
\end{align}
The primary constraint for the MAS is that the specified demand D$_t$ per period must be met. This is taken into account by $g(z)$ in \autoref{eq:4}, which calculates the operating point $\text{Op}_{t}$ of an electrolysis-PEA to meet demand D$_t$. Hence, PEA-agent $n$ aggregates production quantities from all PEA-agents $j \in \mathcal{N}$ within the MAS and determines its own production quantity mH2$_{n,t}$ to minimize the deviation from demand D$_t$.
\vspace{-0.1cm}
\begin{align}
 g(z) &= \sum \limits_{j \in \mathcal{N}\setminus n} \text{mH2}_{j,t} + \text{mH2}_{n,t}(z) - \text{D}_t\label{eq:4}
\end{align}
\autoref{eq:5} ensures that the operating point of an electrolysis-PEA $\text{Op}_{t}$ is within the lower (Op$_\text{min}$) and upper (Op$_\text{max}$) operating limits specified in percent and must be met in both \autoref{eq:3}, \ref{eq:6}. 
\vspace{-0.1cm}
\begin{align}
    \text{Op}_{\text{min}} \leq \text{Op}_{t}(x) \leq \text{Op}_{\text{max}}\label{eq:5}
\end{align}
As detailed in \autoref{eq:6}, a quadratic approximation is used for the hydrogen production rate, following \cite{RWK23, VMZ21}, with production parameters $\alpha, \beta$, $\gamma$, and is applied to \autoref{eq:2} and \ref{eq:4}. Multiplication by $x_{\text{operation},t}$ ensures that the electrolysis-PEA can only produce hydrogen in the operating state. 
\begin{align}
   \text{mH2}_{t} &=  \left(\alpha \cdot \text{Op}_{t}^2(x) + \beta \cdot \text{Op}_{t}(x) + \gamma\right) \cdot x_{\text{operation},t} \label{eq:6}
\end{align}
\autoref{eq:xxxx} ensures that the electrolysis-PEA can only adopt one state $s$ per period (idle or production). Here, $x_{s,t}$ is a binary variable for the respective state: operating  (1 operating, 0 otherwise), and idle (1 idle, 0 otherwise) \cite{WRF23b}. To capture the transition from idle to operation, a holding time $t_h$ is introduced, signifying the duration required for an electrolysis-PEA to shift from the idle state in period $t$ to operation in period $t+t_h$.
\vspace{-0.3cm}
\begin{align}
    \sum \limits_{s \in \mathcal{S}} x_{s,t} = 1 \label{eq:xxxx} 
\end{align}
\autoref{eq:xyz} ensures that the binary variable $Y_t$ is set to 1 when the electrolysis-PEA is started.
\vspace{-0.2cm}
\begin{align}
    x_{\text{operation},t} + x_{\text{idle},t-1} -1 \leq Y_t\label{eq:xyz}
\end{align}
\autoref{eq:minKopplung} and \ref{eq:maxKopplung} couple the state variables with the lower or upper operating limit per state \cite{WRF23b}.
\vspace{-0.1cm}
\begin{align}
    \text{Op}_{t}(x) &\geq \sum \limits_{s \in \mathcal{S}} \text{Op}_{\text{min},s} \cdot x_{s,t}\label{eq:minKopplung}\\
    \text{Op}_{t}(x) &\leq \sum \limits_{s \in \mathcal{S}} \text{Op}_{\text{max},s} \cdot x_{s,t}\label{eq:maxKopplung}
\end{align}
%
In \autoref{eq:7}, the transpose of the vector $y$ denoted as $y^T$ serves as a means to couple the local decision variables, represented by $x$ and $z$, across various PEA-agents. The objective is to maintain consistency in decision-making among these agents. The coupling is facilitated by a Lagrange multiplier $\lambda_t$. This multiplier plays a crucial role in compensating for any disparities that may arise between the local perspectives ($x$) of individual agents and the global perspective ($z$), ensuring that the decisions align harmoniously \cite{Boy10}. The local perspective ($x$) defines the mLCOH-optimal utilization of an electrolysis-PEA, while the global perspective ($z$) represents the utilization of an electrolysis-PEA to meet the total demand. 
\begin{align}
 y^T(Ax + Bz - c) = \lambda_t(x-z)\label{eq:7}
\end{align}
The ADMM algorithm runs through $k$ iterative cycles with the goal of converging to a solution that simultaneously minimizes mLCOH and meets demand D$_t$ in period $t$:
\begin{align}
x^{k+1} &=  \argmin_x \; \mathcal{L}_\rho (x,z^k,y^k )\label{eq:8}\\
z^{k+1} &= \argmin_z \; \mathcal{L}_\rho(x^{k+1}, z, y^k )\label{eq:9}\\
y^{k+1} &= y^k+p(Ax^{k+1}+Bz^{k+1}-c)\label{eq:10}
\end{align}

This iterative approach minimizes one variable at a time while holding the others constant, gradually guiding the solution to a point where both the objective function and the constraints are simultaneously minimized, resulting in an optimal solution to the problem \cite{Boy10}. Initially, \autoref{eq:8} minimizes the Lagrangian function with respect to $x$, while $z$ and $y$ are held constant, and employs the values of the previous iteration \cite{Boy10}. Regarding the decentralized scheduling model, each PEA-agent determines the mLCOH-optimal setpoint of the associated electrolysis-PEA, independent of other constraints.

Then, \autoref{eq:9} minimizes the Lagrangian function with respect to $z$ while holding $x$ and $y$ constant, considering the updated $x$ value from the current iteration. In this step, a PEA-agent determines the operating point that minimizes the deviation from the demand, as defined in \autoref{eq:4}.

The subsequent step incorporates the dual update, as outlined in \autoref{eq:10}. This update process involves the adjustment of the Lagrange multiplier $\lambda_t$ (\autoref{eq:7}). In the dual update, two primary factors are considered. Firstly, the current demand deviation (planned production minus demand in a given period) is factored in to control the rate of convergence. When a higher deviation exists, it leads to an increase in $\lambda_t$, thus expediting the convergence process. Conversely, lower deviations require a less substantial adjustment of $\lambda_t$. Secondly, the current gradient of the mLCOH-function (see \autoref{subsec:PreliminaryWork}) is incorporated. This step ensures that PEA-agents adjust their operating points to effectively reduce costs or minimize cost increases. Moreover, it preserves the autonomy of the PEA-agents, allowing them to decide independently on the change of their operating points.

The \autoref{eq:8}-\ref{eq:10} are performed iteratively until the criterion in \autoref{eq:11}, an allowable deviation to demand $\varepsilon$, is reached:
\begin{align}
    {|x}^{k+1}-x^{k}\left|+{|z}^{k+1}-z^{k}\right|+\left|y^{k+1}-y^{k}\right|<\varepsilon\label{eq:11}
\end{align}

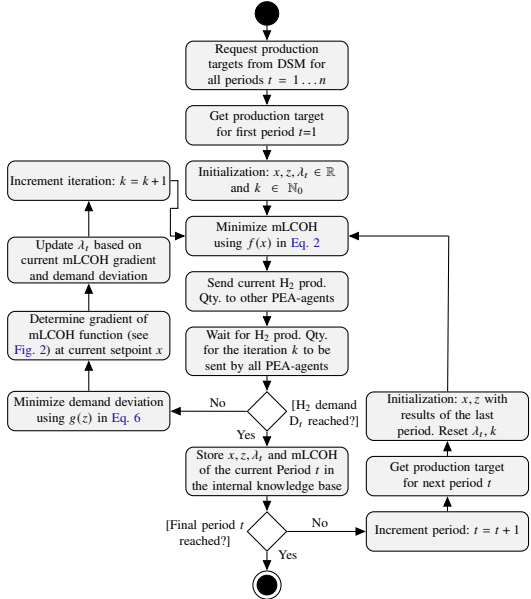
\begin{figure}[h]
\centering
\resizebox{7cm}{!}{\begin{tikzpicture}[auto, thick, >=triangle 45]

	\draw
        node at (3,0) [start](Initial){d}
        node [process, below=of Initial](ProductionTarget){Request production targets from DSM for all periods $t=1 \dots n$}
        node [process, below=of ProductionTarget](DemandFirstPeriod){Get production target\\for first period $t$=1}
        node [process, below=of DemandFirstPeriod](FirstInitialization){Initialization: $x, z, \lambda_t \in \mathbb{R}$ and $k \in \mathbb{N}_0$}
        node [process, below=of FirstInitialization](MinimizeX){Minimize mLCOH using $f(x)$ in \autoref{eq:2}}
        node [process, below=of MinimizeX](BroadcastData){Send current H$_2$ prod. Qty. to other PEA-agents}
        node [process, below=of BroadcastData](WaitForMessages){Wait for H$_2$ prod. Qty. for the iteration $k$ to be sent by all PEA-agents}
        node [decision, below=of WaitForMessages](DecisionDemandReached){\textcolor{white}{dd}}

        node [process, left=of WaitForMessages, yshift = -1.6cm](MinimizeZ){Minimize demand deviation using $g(z)$ in \autoref{eq:4}}
        node [process, left=of WaitForMessages, yshift=4.2mm](DetermineGradient){Determine gradient of mLCOH function (see \autoref{fig:mLCOHGradient}) at current setpoint $x$}
        node [process, left=of BroadcastData, yshift=8.4mm](DualUpdate){Update $\lambda_t$ based on current mLCOH gradient and demand deviation}
        node [process, left=of FirstInitialization](IncrementIteration){Increment iteration: $k = k+1$}

        node [process, below=of DecisionDemandReached](ExtractResult){Store $x,z, \lambda_t$ and mLCOH of the current Period $t$ in the internal knowledge base}
        node [decision, below=of ExtractResult](DecisionFinalPeriod){\textcolor{white}{dd}}
        node [process, right=of DecisionFinalPeriod, xshift = 1.7cm](IncrementPeriod){Increment period: $t = t +1$}
        node [process, above=of IncrementPeriod](DemandNextPeriod){Get production target\\for next period $t$}
        node [process, above=of DemandNextPeriod](InitializationNextPeriod){Initialization: $x, z$  with results of the last period. Reset $\lambda_t, k$}

        node [coordinate, right=of MinimizeX, xshift = 1.68cm] (InvisibleNode1) {}

        node[end, below=of DecisionFinalPeriod] (end) {dd}
        node[endcenter] at (end.center) {}

        node[below right = -0.7cm and 0.2cm of DecisionDemandReached, align=center] {[H$_2$ demand\\ D$_t$ reached?]}
        node[below left = 0.1cm and -0.1cm of DecisionDemandReached, align=center] {Yes}
        node[below left = -0.75cm and 0.7cm of DecisionDemandReached, align=center] {No}

        node[below left = -0.7cm and 0.2cm of DecisionFinalPeriod, align=center] {[Final period $t$\\reached?]}
        node[below right = -0.75cm and 0.8cm of DecisionFinalPeriod, align=center] {No}
        node[below right = 0.1cm and -0.1cm of DecisionFinalPeriod, align=center] {Yes};

        \draw[->](Initial) -- (ProductionTarget);
        \draw[->](ProductionTarget) -- (DemandFirstPeriod);
        \draw[->](DemandFirstPeriod) -- (FirstInitialization);
        \draw[->](FirstInitialization) -- (MinimizeX);
        \draw[->](MinimizeX) -- (BroadcastData);
        \draw[->](BroadcastData) -- (WaitForMessages);
        \draw[->](WaitForMessages) -- (DecisionDemandReached);
        \draw[->](DecisionDemandReached) -- (MinimizeZ);
        \draw[->](MinimizeZ) -- (DetermineGradient);
        \draw[->](DetermineGradient) -- (DualUpdate);
        \draw[->](DualUpdate) -- (IncrementIteration);
       \draw[->] (IncrementIteration.east) -- ++(0.2,0) |- ($(IncrementIteration.east)!0.5!(MinimizeX.west)$) - ++(-0.2,0) |- (MinimizeX.west);
        \draw[->](DecisionDemandReached) -- (ExtractResult);
        \draw[->](ExtractResult) -- (DecisionFinalPeriod);
        \draw[->](DecisionFinalPeriod) -- (IncrementPeriod);
        \draw[->](DecisionFinalPeriod) -- (end);
        \draw[->](IncrementPeriod) -- (DemandNextPeriod);
        \draw[->](DemandNextPeriod) -- (InitializationNextPeriod);
        \draw[-](InitializationNextPeriod) -- (InvisibleNode1);
        \draw[->](InvisibleNode1) -- (MinimizeX);

\end{tikzpicture}}
\vspace{-0.2cm}
\captionsetup{font=small}
\caption{Activity diagram of the decentralized scheduling}
\label{fig:ActivityDiagram}
\end{figure}

The activity diagram in \autoref{fig:ActivityDiagram} visually outlines the steps performed by a PEA-agent to implement the ADMM scheduling model. It is assumed that PEA-agents are mutually aware within the MAS.
The first process step consists of a PEA-agent acquiring production goals for the planning horizon from the DSM, as demonstrated in \autoref{tab:productiongoals}. The PEA-agent then determines the production target for the first period. 

Then, the PEA-agent initializes the variables $x$ and $z$ with randomized values within the upper and lower operating limits, sets $\lambda_t$ and $k$ to 0. The PEA-agent then performs the minimization with respect to $x$ and mLCOH, respectively. The resulting production quantity, which reflects the optimized operating point, is communicated to the other PEA-agents. 

The PEA-agent waits for production quantities for the current iteration $k$ from all other PEA-agents. If no messages are received within a designated timeframe, a PEA-agent is considered inactive, and the production quantity is set to zero. The PEA-agent then checks if the demand has been satisfied and if a tolerable deviation between $x$ and $z$ has been achieved.\\
In case of an unmet demand, the PEA-agent computes the operating point that minimizes the deviation from demand. Additionally, the agent calculates the gradient of the mLCOH function at the current operating point $x$ and updates $\lambda_t$. This approach ensures that each PEA-agent adjusts its operating point proportionately to the gradient. The objective is to guarantee that the PEA-agent with the smallest gradient, indicating the least change in costs, appropriately adjusts its operating point. This is subsequently followed by incrementing $k$ and rerunning the ADMM sequence.

Once the PEA-agents satisfy the demand, the variables $x,z, \lambda_t$ and the resulting mLCOH are stored in their internal knowledge base. A check determines if the last period $t$ of the production targets has been reached. In this case, the scheduling is completed. Otherwise, the period $t$ is incremented and the ADMM sequence starts for the following period $t$. To maintain stability during the transition between planning periods $x$ and $z$ are initialized with the results of the previous period, preventing significant load fluctuations. 
\section{Case Study}\label{sec:CaseStudy}
\vspace{-0.1cm}
The ADMM scheduling model was integrated into a MAS using the \textit{Java Agent Development Framework}.\footnote{The implementation is available at \url{https://github.com/ATHenkel/electrolyzerSchedulingMAS}}
\vspace{-0.1cm}
\subsection{Structure and Methodology}\label{subsec:CaseStudyStructure}
In order to evaluate the ADMM-model described in \autoref{sec:SchedulingModel}, the electrolysis-PEAs of the \textit{Process2Order} (P2O) laboratory \cite{VPK+23} are considered, which are three \textit{Anion Exchange Membrane} (AEM) electrolysers of the type EL 4 by \textit{Enapter}. 

\begin{figure}[ht]
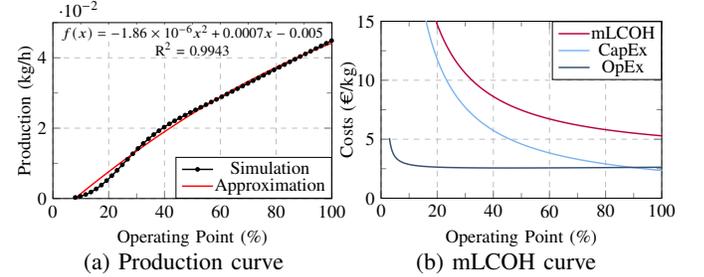

\centering
\begin{subfigure}{.252\textwidth}
  \centering
  \resizebox{4.6cm}{!}{\input{Figures/ProductioncurveAEM}}
  \vspace{-0.65cm}
  \caption{\small Production curve}
  \label{subfig:productioncurveAEM}
\end{subfigure}%
\hspace{-0.4cm}
\begin{subfigure}{.252\textwidth}
  \centering
  \resizebox{4.7cm}{!}{\input{Figures/mLCOH_Enapter}}
  \vspace{-0.65cm}
  \caption{\small mLCOH curve}
  \label{subfig:mLCOHEnapter}
\end{subfigure}
\vspace{-0.6cm}
\captionsetup{font=small}
\caption{(a) Simulated and approximated production curve at different operating points (b) OpEx and CapEx contributions to mLCOH for an AEM EL4 electrolyzer at varying operating points with electricity cost of 0.05€/kWh} 
\end{figure}
\vspace{-0.15cm}
Simulation models\footnote{The authors thank the P2O-lab for providing the simulation models.} are available for these electrolysis-PEAs \cite{VGM+23}, and in \autoref{subfig:productioncurveAEM} the production curve resulting from simulating an electrolyzer as a function of the operation point is depicted, alongside the quadratic approximation of the production curve. The simulation was performed with a lower operating limit Op$_\text{min,operation}$ = 8 \%, an upper operating limit Op$_\text{max, operation}$ = 100\%, and a capacity P$_\text{el}=2.4$kW.
Notably, this quadratic approximation closely aligns with the values of the simulation, as indicated by a coefficient of determination $\mathrm{R}^{2}$ of 0.9943. 

In a study by \textit{Enapter}, financial parameters and the resulting LCOH were published \cite{Ena22}. The former are adopted in this paper and are listed in \autoref{tab:ParameterCaseStudy}. Based on \cite{VMZ21}, the start-up costs were assumed to be 0.12 € per start-up. The mLCOH curve resulting from these financial parameters as a function of power is depicted in \autoref{subfig:mLCOHEnapter}. The AEM electrolyzer, characterized by a high CapEx/kW ratio of 3333€/kW, exhibits optimal mLCOH operation when operated at full load. In contrast, the PEM electrolyzer in \autoref{fig:mLCOHGradient} with a lower CapEx/kW of 820€/kW demonstrates economic feasibility even at lower loads \cite{GVE+22, BRP22}.

\vspace{0.1cm}
\begin{table}[h]
  \centering
  \captionsetup{font=small}
  \caption{Financial parameters (Adopted from \cite{Ena22})}
  \vspace{-0.1cm}
  \renewcommand{\arraystretch}{1.1} %
    \begin{tabular}{|l|c|c|l|}
    \hline
    \textbf{Parameter} & \multicolumn{1}{l|}{\textbf{Symbol}} & \textbf{Value}  & \textbf{Unit} \\
    \hline
    CapEx$_0$ & - & 8000  & € \\
    \hline
    O\&M cost & \textit{OMF} & 1.5   & \% of CapEx$_0$ / year \\
    \hline
    Utilization time & $UT$  & 20    & years \\
    \hline
    Load factor & \textit{LF} & 98    & \% \\
    \hline
    Discount rate & $r$ & 9.73  & \% \\
    \hline
    \end{tabular}%
  \label{tab:ParameterCaseStudy}%
  \vspace{-0.2cm}
\end{table}%

To validate the scheduling model, the evaluation occurs in two steps. First, mLCOH values are compared with results from \cite{Ena22} under nominal load conditions (100\%). In the second phase, dynamic behavior is examined. To emphasize robustness, production targets were randomly generated within the feasible range detailed in \autoref{tab:productiongoals}, considering the aggregated lower and upper operating limits of the three AEM electrolysis-PEAs.
In addition, the scale-up and malfunction of an electrolysis-PEA were simulated to evaluate the low-effort scalability (R2) and robustness (R3) of the scheduling model.

\vspace{0.1cm}
\begin{table}[h]
  \centering
  \captionsetup{font=small}
  \caption{Production targets for 12 periods at 15-minute intervals}
  \vspace{-0.1cm}
  \renewcommand{\arraystretch}{1.1} %
  \newcolumntype{M}[1]{>{\centering\arraybackslash}b{#1}} 
    \begin{tabular}{|M{0.1cm}|M{1.1cm}|M{1.9cm}|M{0.1cm}|M{1.1cm}|M{1.9cm}|}
    \hline
\textbf{$t$} & \centering \textbf{D$_t$}(kg/h) & \textbf{C${_{\text{E},t}}$}(€/MWh) & \textbf{$t$} & \textbf{D$_t$} (kg/h) & \textbf{C${_{\text{E},t}}$}(€/MWh)\\
    \hline
    1     & 0.1320 & 19.48 & 7     & 0.0881 & 10.01 \\
    \hline
    2     & 0.0781 & 55.54 & 8     & 0.0989 & 103.94 \\
    \hline
    3     & 0.0669 & 54.09 & 9     & 0.0271 & 105.42 \\
    \hline
    4     & 0.1262 & 33.42 & \!10    & 0.0758 & 116.16 \\
    \hline
    5     & 0.0413 & 49.60 & \!11    & 0.0344 & 109.01 \\
    \hline
    6     & 0.0469 & 63.95 & \!12    & 0.0812 & 105.55 \\
    \hline
    \end{tabular}%
  \label{tab:productiongoals}%
\end{table}%


\subsection{Results and Discussion}\label{subsec:CaseStudyResults}
In \autoref{tab:LCOHComparison}, mLCOH values taken from the ADMM model are compared with LCOH values from \cite{Ena22} under an electricity price of 0.05€/kWh and nominal load (100\%). Please note that the OpEx result from the electricity costs.
\vspace{0.1cm}
\begin{table}[h]
  \centering
  \captionsetup{font=small}
  \caption{LCOH results of \textit{Enapter} \cite{Ena22} and the scheduling model at nominal load (100\%) and an electricity price of 0.05€/kWh}
  \vspace{-0.1cm}
  \renewcommand{\arraystretch}{1.1} %
    \begin{tabular}{|l|c|c|}
    \hline
    \textbf{Cost-Factor} & \textbf{\textit{Enapter} \cite{Ena22}} & \textbf{ADMM-Model} \\
    \hline
    CapEx (€/kg) & 2.39  & 2.39 \\
    \hline
    OpEx (€/kg) & 2.67  & 2.67 \\
    \hline
    O\&M (€/kg) & 0.31  & 0.31 \\
    \hline
    LCOH (€/kg) & 5.37  & 5.37 \\
    \hline
    \end{tabular}%
  \label{tab:LCOHComparison}%
\end{table}%
The results demonstrate that the ADMM model accurately calculates the mLCOH. 
This comparison highlights that considering mLCOH, which focuses on a single period \cite{GVE+22}, does not significantly differ from the more comprehensive LCOH, which spans the entire utilization time of an electrolyzer \cite{BRP22}.

\begin{figure}[h]
\centering
\begin{subfigure}{.246\textwidth}
  \centering
  \resizebox{0.97\textwidth}{!}{\input{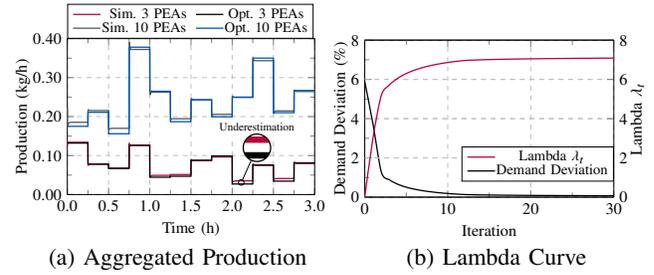}}
  \vspace{-0.15cm}
  \caption{\small Aggregated Production}
  \label{subfig:SimulationResults}
\end{subfigure}%
\hspace{-0.5cm} 
\begin{subfigure}{.263\textwidth}
  \centering
  \resizebox{0.91\textwidth}{!}{\begin{tikzpicture}[spy using outlines={circle, magnification=6, connect spies}]
\begin{axis} [ /pgf/number format/.cd,
        1000 sep={.},
        height=5cm, 
        width=7cm,
        xmin=0, xmax=30,
        minor x tick num=1,
        minor y tick num=1,
        ymin=0, ymax=8, 
        xlabel={Iteration},
        ylabel={Demand Deviation (\%)},
        ylabel style={yshift=-5pt},
        ymajorgrids=true,
        xmajorgrids=true,
        grid style=dashed, 
        yticklabel pos=left ,axis y line*=left, 
        x tick label style={
            /pgf/number format/.cd,
            fixed,
            fixed zerofill,
            precision=0,
        },
    ]

\addplot[smooth, mark=,color=black, thick] plot coordinates {
(0,5.89140928964481)
(1,3.71309495421275)
(2,1.2747544864425)
(3,0.87669412088249)
(4,0.65363948238753)
(5,0.50825886079955)
(6,0.40523968467458)
(7,0.329610371299589)
(8,0.27043371922659)
(9,0.22361288362661)
(10,0.187790842306619)
(11,0.15746778567462)
(12,0.134028736831619)
(13,0.122996304799619)
(14,0.11472101838764)
(15,0.10920370249964)
(16,0.103686019987639)
(17,0.0995475175066401)
(18,0.0954088087996394)
(19,0.0912698938666398)
(20,0.0885105026746394)
(21,0.0857510198266395)
(22,0.0829914453226388)
(23,0.08161162369964)
(24,0.0788519117116387)
(25,0.077472021346639)
(26,0.0760921080676394)
(27,0.0733322127676406)
(28,0.0719522307466386)
(29,0.0705722258116395)
(30,0.0691921979626392)
(31,0.067812147199639)
(32,0.066432073522639)
(33,0.0650519769316391)
(34,0.0636718574266393)
(35,0.0636718574266393)
(36,0.0622917150076396)
(37,0.0609115496746387)
(38,0.0595313614276394)
(39,0.0595313614276394)
(40,0.0581511502666401)
(41,0.0567709161916397)
(42,0.0567709161916397)
(43,0.0553906592026393)
(44,0.0540103792996405)
(45,0.0540103792996405)
(46,0.0526300764826404)
(47,0.0526300764826404)
(48,0.0512497507516391)
(49,0.0512497507516391)

};
\label{demanddev}
\end{axis}

\begin{axis} [ /pgf/number format/.cd,
        1000 sep={.},
          legend style={
            at={(1,0.1)},
            anchor=south east,
            nodes={inner sep=0.5pt},
            font=\small, 
        },
        height=5cm, 
        width=7cm,
        xmin=0, xmax=30,
        ymin=0, ymax=8, 
        ylabel={Lambda $\lambda_t$},
        ymajorgrids=true,
        xmajorgrids=true,
        ylabel style={yshift=5pt},
        grid style=dashed, 
       yticklabel pos=right ,axis y line*=right, 
       xlabel={}, xtick=\empty,
        x tick label style={
            /pgf/number format/.cd,
            fixed,
            fixed zerofill,
            precision=0,
        },
    ]
\addplot[smooth, mark=, color=hsurot, thick] plot coordinates {
(0,0)
(1,3.03361843446097)
(2,5.19239001899316)
(3,5.73096463571714)
(4,6.06632851394265)
(5,6.30002000875159)
(6,6.47306202831952)
(7,6.60580057156499)
(8,6.71065750499307)
(9,6.79447029579385)
(10,6.86241947175371)
(11,6.91851697861918)
(12,6.96141756854407)
(13,6.98309284309173)
(14,6.99865336247054)
(15,7.01072534464434)
(16,7.02088856965508)
(17,7.02942279351626)
(18,7.03689511029137)
(19,7.043426245342)
(20,7.04912391241753)
(21,7.05431996663617)
(22,7.0590539257646)
(23,7.06336238235283)
(24,7.06747095303025)
(25,7.07120387631521)
(26,7.07476043036859)
(27,7.0781478646987)
(28,7.08121771057555)
(29,7.0841385638647)
(30,7.08691662565634)
(31,7.08955785845999)
(32,7.09206799495323)
(33,7.09445254641971)
(34,7.09671681088753)
(35,7.09886588097814)
(36,7.10101495106876)
(37,7.10305372156663)
(38,7.10498689671918)
(39,7.10681899727656)
(40,7.10865109783394)
(41,7.11038646783753)
(42,7.11202928116603)
(43,7.11367209449453)
(44,7.11522636117601)
(45,7.1166959334741)
(46,7.11816550577219)
(47,7.11955408405432)
(48,7.12094266233644)
(49,7.12225380073164)
};
\addlegendentry{Lambda $\lambda_{t}$}
\addlegendimage{/pgfplots/legend to name=demanddev, mark=,color=black, thick}
\addlegendentry{Demand Deviation}
\end{axis}

\end{tikzpicture}}
  \vspace{-0.15cm}
  \caption{\small Lambda Curve}
  \label{subfig:lambdaCurve}
\end{subfigure}
\vspace{-0.6cm}
\captionsetup{font=small}
\caption{(a) Aggregated optimized schedule (Opt.) and simulation data (Sim.) over 12 periods (b) Demand deviation and lambda for PEA-agent 1 in period $t=10$} 
\vspace{-2em}
\end{figure}

The optimized scheduling results corresponding to the demand in \autoref{tab:productiongoals} are illustrated in \autoref{subfig:SimulationResults}. The optimized operating points were communicated to the simulation using a standardized MTP interface \cite{2658}. Monitoring was similarly integrated via the same interface (R6).

The graph illustrates that the scheduling results closely mirror the simulation, as evidenced by a normalized \textit{Root Mean Squared Error} (nRMSE) of 3.3\%. However, in the realm of lower utilization rates (30-60\%), the approximation (refer to \autoref{subfig:productioncurveAEM}) underestimates the production curve, consequently leading to higher deviations in the scheduling outcomes in \autoref{subfig:SimulationResults}. It is worth noting that in this work, homogeneity among the electrolysis-PEAs results in equal utilization, hence the aggregated schedule showcases the overall behavior of the electrolysis system.

In another scenario, the number of electrolysis-PEAs was increased from 3 to 10. 
Production targets are generated randomly following the pattern outlined in \autoref{tab:productiongoals} and adjusted in proportion to the expanded total production capacity. The resulting optimized scheduling and the values measured during the simulation are depicted in \autoref{subfig:SimulationResults}.
The decentralized scheduling model showcases scalability, accommodating an increase in the number of electrolysis-PEAs from 3 to 10. Importantly, no manual configuration effort was required. Additional PEA-agents were instantiated based on the number of electrolysis-PEAs and cooperatively performed the scheduling.
These observations align with results from \cite{EHS15, SANA22}, affirming that a decentralized approach can be adapted with minimal effort, without the need for extra variables, constraints, or cost functions. Such characteristics make this approach particularly suitable for modular electrolysis plants automated with MTP, where low-effort expansions and adaptability (R2) are considered crucial \cite{BBE+22, LKL+23c}.

In \autoref{subfig:lambdaCurve}, the lambda curve ($\lambda_t$) for PEA-agent 1 in the period $t = 10$ is shown as a function of the iterations. As described in \autoref{sec:SchedulingModel}, the Lagrange multiplier $\lambda_t$ is employed to couple $x$ and $z$, aiming to align the local and global perspectives. This coupling mechanism is integral to the dual update process, which, as outlined in \autoref{sec:SchedulingModel}, considers the prevailing deviation from the demand to regulate the convergence speed.
Early iterations reveal a substantial deviation (approx. 6\%) resulting in a sharp increase in $\lambda_t$ up to iteration 5. Subsequently, the demand deviation narrows to less than 1\%, yielding a more gradual increment in $\lambda_t$ per iteration. This observation underscores the ability of the dual update, as described in \autoref{sec:SchedulingModel}, to regulate the rate of convergence by incorporating the demand deviation, preventing oscillations and ensuring a stable convergence process.

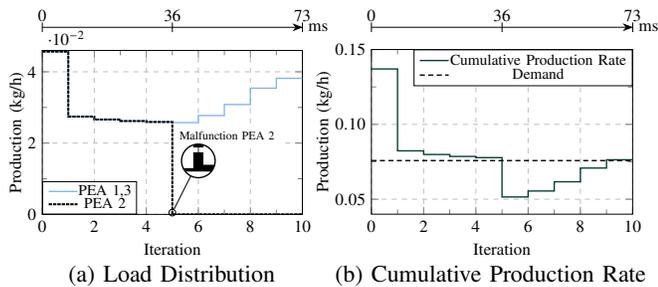
\begin{figure}[ht]
\centering
\begin{subfigure}{.246\textwidth}
  \centering
  \resizebox{\textwidth}{!}{\begin{tikzpicture}[spy using outlines={circle, magnification=5, connect spies}]
\begin{axis} [ /pgf/number format/.cd,
        1000 sep={.},
        legend style={
            at={(0,0)},
            anchor=south west,
            nodes={inner sep=0.pt},
            font=\small, 
        },
        height=5cm, 
        width=7cm,
        xmin=0, xmax=10,
        minor x tick num=1,
        minor y tick num=1,
        ymin=0, ymax=0.046, 
        xlabel={Iteration},
        ylabel={Production (kg/h)},
        ylabel style={yshift=-5pt},
        ymajorgrids=true,
        xmajorgrids=true,
        grid style=dashed, 
        x tick label style={
            /pgf/number format/.cd,
            fixed,
            fixed zerofill,
            precision=0,
        },
    ]

\addplot[const plot, mark=,color=babyblau, thick] plot coordinates {
(0,0.045671364)
(1,0.027439105)
(2,0.026622552)
(3,0.026190362)
(4,0.025927961)
(5,0.025752889)
(6,0.027736681)
(7,0.030825915)
(8,0.035404854)
(9,0.038177336)
(10,0.038177336)
};

\addplot[const plot, mark=,color=black, dash pattern=on .7mm off .2mm, very thick] plot coordinates {
(0,0.045671364)
(1,0.027439105)
(2,0.026622552)
(3,0.026190362)
(4,0.025927961)
(5,0)
(6,0)
(7,0)
(8,0)
(9,0)
(10,0)
};


\coordinate (spypoint1) at (axis cs: 5,0.0005);
\coordinate (magnifyglass1) at (axis cs:6,0.015);
\node[anchor=west, fill=white] at (axis cs:5.05,0.022) {\textcolor{black!100}{\scriptsize Malfunction PEA 2}};

\addlegendentry{PEA 1,3}
\addlegendentry{PEA 2}

\end{axis}
\spy [black, size=.7cm] on (spypoint1) in node[fill=white] at (magnifyglass1);

\draw[-{Stealth[length=2mm]}] (0, 3.9) node[left, above] {0} -- (5.42, 3.9) node[right] {ms}
node[midway, above] {36} 
node[above] {73};
\draw (0, 3.8) -- (0, 4); 
\draw (2.71, 3.8) -- (2.71, 4); 
\draw (5.42, 3.8) -- (5.42, 4); 

\end{tikzpicture}}
  \vspace{-0.6cm}
  \caption{\small Load Distribution}
  \label{subfig:ElectrolyzerFailure}
\end{subfigure}%
\hspace{-0.5cm} 
\begin{subfigure}{.263\textwidth}
  \centering
  \resizebox{\textwidth}{!}{\begin{tikzpicture}[spy using outlines={circle, magnification=6, connect spies}]
\begin{axis} [ /pgf/number format/.cd,
        1000 sep={.},
        legend style={
            at={(1,1)},
            anchor=north east,
            nodes={inner sep=0.5pt},
            font=\small, 
        },
        height=5cm, 
        width=7cm,
        xmin=0, xmax=10,
        minor x tick num=1,
        minor y tick num=1,
        ymin=0.04, ymax=0.15, 
        xlabel={Iteration},
        ylabel={Production (kg/h)},
        ylabel style={yshift=-5pt},
        ymajorgrids=true,
        xmajorgrids=true,
        grid style=dashed, 
        x tick label style={
            /pgf/number format/.cd,
            fixed,
            fixed zerofill,
            precision=0,
        },
         y tick label style={
            /pgf/number format/.cd,
            fixed,
            fixed zerofill,
            precision=2,
        },
    ]

\addplot[const plot, mark=,color=britishracinggreen, thick] plot coordinates {
(0,0.137014093)
(1,0.082317314)
(2,0.079867656)
(3,0.078571086)
(4,0.077783883)
(5,0.051505778)
(6,0.055473362)
(7,0.06165183)
(8,0.070809708)
(9,0.076354673)
(10,0.076)

};

\addplot[const plot, mark=,color=black,densely dashed, thick] plot coordinates {
(0,0.0758)
(1,0.0758)
(2,0.0758)
(3,0.0758)
(4,0.0758)
(5,0.0758)
(6,0.0758)
(7,0.0758)
(8,0.0758)
(9,0.0758)
(10,0.0758)
};

\addlegendentry{Cumulative Production Rate}
\addlegendentry{Demand}
\end{axis}

\draw[-{Stealth[length=2mm]}] (0, 3.88) node[left, above] {0}  -- (5.42, 3.88) node[right] {ms}
node[midway, above] {36} 
node[above] {73};

\draw (0, 3.78) -- (0, 3.98); 
\draw (2.71, 3.78) -- (2.71, 3.98); 
\draw (5.42, 3.78) -- (5.42, 3.98); 

\end{tikzpicture}}
  \vspace{-0.6cm}
  \caption{\small Cumulative Production Rate}
  \label{subfig:ElectrolyzerFailureCumulativeProduction}
\end{subfigure}
\vspace{-0.6cm}
\captionsetup{font=small}
\caption{Simulated malfunction of electrolysis-PEA 2 in period $t$=10 and iteration $k$=5. (a) Load distribution and malfunction (b) Cumulative production rate and demand}
\vspace{-1.25em}
\end{figure}

In \autoref{subfig:ElectrolyzerFailure} and \ref{subfig:ElectrolyzerFailureCumulativeProduction}  the response of the model to an electrolysis-PEA malfunction is illustrated. A malfunction of electrolysis-PEA 2 (corresponding to PEA-agent 2) in period 10 and iteration 5 was simulated. It is evident that the PEA-agents within the next iteration can recognize the malfunction of the electrolysis-PEA, as they exchange the respective production quantities with each other (see \autoref{fig:ActivityDiagram}). 
Accordingly, \autoref{eq:4} becomes operative, as the malfunction leads to a higher discrepancy of production and demand.

Consequently, the remaining operational PEA-agents 1 and 3 increase the production quantity to meet the demand. The entire process, from the disruption notification to the readjustment of production quantities to meet demand, takes about 4 iterations. Using the applied hardware (AMD Ryzen 7 PRO, 32GB RAM) the actual time for this rescheduling was less than 40ms (refer to \autoref{subfig:ElectrolyzerFailure} - \ref{subfig:ElectrolyzerFailureCumulativeProduction}, 2nd x-axis, iterations 5 to 10). This emphasizes the fact that the ADMM model itself is very robust as well as fast solving and that no separate disturbance management routines need to be executed to cope with electrolysis-PEA or PEA-agent malfunctions.
\section{Conclusion and Future Work}\label{sec:conclusion}
This paper has presented a decentralized ADMM-based scheduling model using MAS to efficiently optimize the operation of modular electrolysis plants. The model aligns hydrogen production with fluctuating demand, minimizing mLCOH and ensuring adaptability to operational challenges.

A case study displayed the accuracy of the model in calculating mLCOH values and its alignment with existing parameters. In particular, the model exhibited responsiveness to dynamic changes, such as electrolysis-PEA malfunctions, and scale-up scenarios, efficiently meeting demand while minimizing hydrogen production costs.

The ADMM-based scheduling model emerges as a promising solution for optimizing the operation of modular electrolysis plants. Unlike existing approaches, the decentralized nature of the ADMM model aligns seamlessly with the decentralized structure of modular electrolysis plants, marked by numerous electrolysis-PEAs \cite{LKL+23c}. This alignment ensures a resilient optimization strategy, capable of addressing deviations, such as electrolysis-PEA malfunctions, and responding dynamically to real-world operational complexities.
Future research should focus on real-world operations in the P2O-Lab \cite{VPK+23} to investigate rescheduling  during electrolysis-PEA malfunctions. Additionally, investigating the computational effort for large, heterogenously configured electrolysis plants with numerous electrolysis-PEAs represents a valuable direction for further research.

\bibliographystyle{IEEEtranN}

\begin{thebibliography}{30}
\providecommand{\natexlab}[1]{#1}
\providecommand{\url}[1]{#1}
\csname url@samestyle\endcsname
\providecommand{\newblock}{\relax}
\providecommand{\bibinfo}[2]{#2}
\providecommand{\BIBentrySTDinterwordspacing}{\spaceskip=0pt\relax}
\providecommand{\BIBentryALTinterwordstretchfactor}{4}
\providecommand{\BIBentryALTinterwordspacing}{\spaceskip=\fontdimen2\font plus
\BIBentryALTinterwordstretchfactor\fontdimen3\font minus
  \fontdimen4\font\relax}
\providecommand{\BIBforeignlanguage}[2]{{%
\expandafter\ifx\csname l@#1\endcsname\relax
\typeout{** WARNING: IEEEtranN.bst: No hyphenation pattern has been}%
\typeout{** loaded for the language `#1'. Using the pattern for}%
\typeout{** the default language instead.}%
\else
\language=\csname l@#1\endcsname
\fi
#2}}
\providecommand{\BIBdecl}{\relax}
\BIBdecl

\bibitem[Odenweller et~al.(2022)Odenweller, Ueckerdt, Nemet, Jensterle, and
  Luderer]{OUN+22}
A.~Odenweller, F.~Ueckerdt, G.~F. Nemet \emph{et~al.}, ``Probabilistic
  feasibility space of scaling up green hydrogen supply,'' \emph{Nature
  Energy}, vol.~7, no.~9, pp. 854--865, 2022.

\bibitem[Bittorf et~al.(2022)Bittorf, Beisswenger, Erdmann, Lorenz, Klose,
  Lange, Urbas, Markaj, and Fay]{BBE+22}
L.~Bittorf, L.~Beisswenger, D.~Erdmann \emph{et~al.}, ``Upcoming domains for
  the mtp and an evaluation of its usability for electrolysis,'' in \emph{2022
  27th ETFA}.\hskip 1em plus 0.5em minus 0.4em\relax IEEE, 2022.

\bibitem[Lange et~al.(2023)Lange, Klose, Lippmann, and Urbas]{LKL+23}
H.~Lange, A.~Klose, W.~Lippmann \emph{et~al.}, ``Technical evaluation of the
  flexibility of water electrolysis systems to increase energy flexibility: A
  review,'' \emph{International Journal of Hydrogen Energy}, 2023.

\bibitem[Lorenz et~al.(2023{\natexlab{a}})Lorenz, Klose, Lange, Kock, and
  Urbas]{LKL+23c}
J.~Lorenz, A.~Klose, H.~Lange \emph{et~al.}, ``Flexible process control for
  scalable electrolysis systems,'' in \emph{2023 ICECET}.\hskip 1em plus 0.5em
  minus 0.4em\relax IEEE, 2023, pp. 1--6.

\bibitem[Nedi{\'c} et~al.(2018)Nedi{\'c}, Olshevsky, and
  Rabbat]{nedic2018network}
A.~Nedi{\'c}, A.~Olshevsky, and M.~G. Rabbat, ``Network topology and
  communication-computation tradeoffs in decentralized optimization,''
  \emph{Proceedings of the IEEE}, vol. 106, no.~5, pp. 953--976, 2018.

\bibitem[Scholz et~al.(2023)Scholz, Henkel, Markaj, and Fay]{SHM+23}
L.~Scholz, V.~Henkel, A.~Markaj \emph{et~al.}, ``Towards a domain-specific
  refinement of process orchestration layers for modular electrolysis plants,''
  in \emph{2023 28th ETFA}.\hskip 1em plus 0.5em minus 0.4em\relax IEEE, 2023.

\bibitem[Huckert et~al.(2024)Huckert, Sidorenko, and Wagner]{HSW24}
J.~L. Huckert, A.~Sidorenko, and A.~Wagner, ``Analysis and assessment of
  multi-agent systems for production planning and control,'' in \emph{Flexible
  Automation and Intelligent Manufacturing: Establishing Bridges for More
  Sustainable Manufacturing Systems}.\hskip 1em plus 0.5em minus 0.4em\relax
  Cham: {Springer Nature Switzerland}, 2024, pp. 687--698.

\bibitem[Reinpold et~al.(2023)Reinpold, Wagner, Gehlhoff, Ramonat, Kilthau,
  Gill, Reif, Henkel, Scholz, and Fay]{RWG+23}
L.~M. Reinpold, L.~P. Wagner, F.~Gehlhoff \emph{et~al.}, ``{Systematic
  Comparison of Software Agents and Digital Twins: Differences, Similarities,
  and Synergies in Industrial Production},'' \emph{{Journal of Intelligent
  Manufacturing}}, 2023.

\bibitem[VDI(October 2019)]{2658}
\emph{Automation engineering of modular systems in the process industry},
  VDI/VDE/NAMUR Std. 2658, October 2019.

\bibitem[Ginsberg et~al.(2022)Ginsberg, Venkatraman, Esposito, and
  Fthenakis]{GVE+22}
M.~J. Ginsberg, M.~Venkatraman, D.~V. Esposito \emph{et~al.}, ``Minimizing the
  cost of hydrogen production through dynamic polymer electrolyte membrane
  electrolyzer operation,'' \emph{Cell Reports Physical Science}, 2022.

\bibitem[Varela et~al.(2021)Varela, Mostafa, and Zondervan]{VMZ21}
C.~Varela, M.~Mostafa, and E.~Zondervan, ``Modeling alkaline water electrolysis
  for power-to-x applications: A scheduling approach,'' \emph{International
  Journal of Hydrogen Energy}, pp. 9303--9313, 2021.

\bibitem[Vincenti et~al.(2022)Vincenti, Cominini, Furlanetto, Sorlini, and
  Valenti]{VCF+22}
F.~Vincenti, P.~Cominini, D.~Furlanetto \emph{et~al.}, ``Optimized size and
  schedule of the power-to-hydrogen system connected to a hydrogen refuelling
  station for waste transportation vehicles in valle camonica,'' \emph{Journal
  of Physics}, vol. 2385, no.~1, p. 012039, 2022.

\bibitem[Flamm et~al.(2021)Flamm, Peter, B{\"u}chi, and Lygeros]{FPB+21}
B.~Flamm, C.~Peter, F.~N. B{\"u}chi \emph{et~al.}, ``Electrolyzer modeling and
  real-time control for optimized production of hydrogen gas,'' \emph{Applied
  Energy}, vol. 281, p. 116031, 2021.

\bibitem[Fang and Liang(2019)]{FaLi19}
R.~Fang and Y.~Liang, ``Control strategy of electrolyzer in a wind-hydrogen
  system considering the constraints of switching times,'' \emph{International
  Journal of Hydrogen Energy}, pp. 25\,104--25\,111, 2019.

\bibitem[Lorenz et~al.(2023{\natexlab{b}})Lorenz, Klose, Lange, Kock, and
  Urbas]{LKL+23b}
J.~Lorenz, A.~Klose, H.~Lange \emph{et~al.}, ``Process control principles for
  scalable electrolysis systems,'' in \emph{2023 IEEE 28th ETFA}.\hskip 1em
  plus 0.5em minus 0.4em\relax IEEE, 2023.

\bibitem[Zhao et~al.(2022)Zhao, Zhu, Tang, Guo, and Sun]{ZZT+22}
Y.~Zhao, Z.~Zhu, S.~Tang \emph{et~al.}, ``Electrolyzer array alternate control
  strategy considering wind power prediction,'' \emph{Energy Reports}, 2022.

\bibitem[Khaligh et~al.(2022)Khaligh, Ghezelbash, Mazidi, Liu, Ryu, and
  Na]{KGM+22}
V.~Khaligh, A.~Ghezelbash, M.~Mazidi \emph{et~al.}, ``A stochastic agent-based
  cooperative scheduling model of a multi-vector microgrid including
  electricity, hydrogen, and gas sectors,'' \emph{Journal of Power Sources},
  2022.

\bibitem[Ma et~al.(2022)Ma, Pei, Deng, Xiao, Yang, and Tang]{MPD+22}
T.~Ma, W.~Pei, W.~Deng \emph{et~al.}, ``A nash bargaining-based cooperative
  planning and operation method for wind-hydrogen-heat multi-agent energy
  system,'' \emph{Energy}, vol. 239, 2022.

\bibitem[Barakat et~al.(2020)Barakat, Tala-Ighil, Gualous, and Hissel]{BTG+20}
R.~Barakat, B.~Tala-Ighil, H.~Gualous \emph{et~al.}, ``Jade-based multi-agent
  decentralized energy management system of a hybrid marine-hydrogen power
  generation system,'' in \emph{ELECTRIMACS}.\hskip 1em plus 0.5em minus
  0.4em\relax {Springer, Cham}, 2020.

\bibitem[Eksin et~al.(2015)Eksin, Hooshmand, and Sharma]{EHS15}
C.~Eksin, A.~Hooshmand, and R.~Sharma, ``A decentralized energy management
  system,'' in \emph{2015 ECC}.\hskip 1em plus 0.5em minus 0.4em\relax IEEE,
  2015, pp. 2260--2267.

\bibitem[Safari and Nasiraghdam(2022)]{SANA22}
A.~Safari and H.~Nasiraghdam, ``Stochastic day-ahead optimal scheduling of
  multimicrogrids: an alternating direction method of multipliers (admm)
  approach,'' \emph{Turkish Journal of Electrical Engineering and Computer
  Sciences}, vol.~30, no.~4, pp. 1354--1369, 2022.

\bibitem[Boyd(2010)]{Boy10}
S.~Boyd, ``Distributed optimization and statistical learning via the
  alternating direction method of multipliers,'' \emph{Foundations and
  Trends{\circledR} in Machine Learning}, vol.~3, no.~1, pp. 1--122, 2010.

\bibitem[Kost et~al.(2013)Kost, Mayer, Thomsen, Hartmann, Senkpiel, Philipps,
  Nold, Lude, Saad, and Schlegl]{KSF+13}
C.~Kost, J.~N. Mayer, J.~Thomsen \emph{et~al.}, ``Levelized cost of
  electricity-renewable energy technologies,'' 2013.

\bibitem[Badgett et~al.(2022)Badgett, Ruth, and Pivovar]{BRP22}
A.~Badgett, M.~Ruth, and B.~Pivovar, ``Economic considerations for hydrogen
  production with a focus on polymer electrolyte membrane electrolysis,'' in
  \emph{Electrochemical Power Sources: Fundamentals, Systems, and
  Applications}.\hskip 1em plus 0.5em minus 0.4em\relax Elsevier, 2022.

\bibitem[Badgett et~al.(2023)Badgett, Brauch, Saha, and Pivovar]{BBS+23}
A.~Badgett, J.~Brauch, P.~Saha \emph{et~al.}, ``Decarbonization of the electric
  power sector and implications for low--cost hydrogen production from water
  electrolysis,'' \emph{Advanced Sustainable Systems}, 2023.

\bibitem[Raheli et~al.(2023)Raheli, Werner, and Kazempour]{RWK23}
E.~Raheli, Y.~Werner, and J.~Kazempour, ``A conic model for electrolyzer
  scheduling,'' \emph{Computers {\&} Chemical Engineering}, vol. 179, 2023.

\bibitem[Wagner et~al.(2023)Wagner, Reinpold, and Fay]{WRF23b}
L.~P. Wagner, L.~M. Reinpold, and A.~Fay, ``{Design Patterns for Optimization
  Models of Flexible Energy Resources},'' in \emph{{2nd IEEE Industrial
  Electronics Society Annual Online Conference (ONCON)}}, online, 2023.

\bibitem[Vogt et~al.(2023)Vogt, Pelzer, Klose, Khadyrov, Lange, Viedt, Urbas,
  and M{\"a}dler]{VPK+23}
L.~Vogt, F.~Pelzer, A.~Klose \emph{et~al.}, ``P2o-lab: A learning factory for
  digitalization and modularization,'' \emph{SSRN Electronic Journal}, 2023.

\bibitem[Viedt et~al.(2023)Viedt, Gopa, M{\"a}dler, and Urbas]{VGM+23}
I.~Viedt, K.~R. Gopa, J.~M{\"a}dler \emph{et~al.}, ``Quality assessment of
  partial models for co-simulation of modular electrolysis plants,'' in
  \emph{33rd ESCAPE}, ser. Computer Aided Chemical Engineering.\hskip 1em plus
  0.5em minus 0.4em\relax Elsevier, 2023.

\bibitem[{Enapter AG}()]{Ena22}
\BIBentryALTinterwordspacing
{Enapter AG}, ``Lcoh assumptions (enapter aem electrolysers).'' [Online].
  Available:
  \url{https://www.enapter.com/app/uploads/2022/07/2-LCOH-Blog-Post_Summary_Rev-2.pdf}
\BIBentrySTDinterwordspacing

\end{thebibliography}
\footnotesize{

}
\end{document}